# Classical-field theory of thermal radiation

**Sergey A. Rashkovskiy**

*Institute for Problems in Mechanics of the Russian Academy of Sciences, Vernadskogo Ave., 101/1, Moscow, 119526, Russia*

*Tomsk State University, 36 Lenina Avenue, Tomsk, 634050, Russia*

*E-mail: rash@ipmnet.ru, Tel. +7 906 0318854*

March 8, 2016

**Abstract** In this paper, using the viewpoint that quantum mechanics can be constructed as a classical field theory without any quantization I build a fully classical theory of thermal radiation. Planck's law for the spectral energy density of thermal radiation and the Einstein A-coefficient for spontaneous emission are derived in the framework of classical field theory without using the concept of "photon". It is shown that the spectral energy density of thermal radiation is apparently not a universal function of frequency, as follows from the Planck's law, but depends weakly on the nature of atoms, while Planck's law is valid only as an approximation in the limit of weak excitation of atoms.

## 1 Introduction

Thermal black body radiation was the first physical phenomenon for whose explanation the idea of the quantization of energy was used (M. Planck, 1900).

As is well known, Planck's idea of quantization of the energy of the oscillator was initially met with scepticism. Many prominent physicists of the time were inclined to consider Planck's proposed method of calculating the mean energy of the oscillator as a mathematical technique (i.e., a trick) with no serious physical meaning.

Serious support for Planck's idea was provided by A. Einstein (1905, 1916), who generalised it to quantization of the energy of the radiation field. Einstein also introduced the concept of stimulated and spontaneous emission after considering the kinetics of the process of "emission-absorption of a quantum by an atom"; on this basis he proposed a new, kinetic derivation of Planck's law.

Einstein's proposed method of derivation of the Planck's law for the spectrum of equilibrium thermal radiation, and particularly the relationship he obtained between the intensities of the absorption and the stimulated emission, as well as between the intensities of spontaneous and stimulated emission, became somewhat of a benchmark that focussed all subsequent researchers on this issue. Additionally, the ability to obtain the so-called Einstein A-coefficient for spontaneous emission began to be seen as convincing evidence of the correctness of any theory of thermal radiation.

Subsequently, the internally consistent quantum theory was built from the disparate facts that energy quanta successfully explained; however, there was a price to pay for it. The price was the rejection of the classic images of waves and particles and the consequent the loss of clarity of the theory: the objects that the theory describes cannot be imagined. However, it turned out that one becomes accustomed to this.

Dirac's quantum radiation theory became the apex of this theory, which later became quantum electrodynamics. It was capable to describe (but not to understand!) the totality of atomic phenomena and phenomena of the interaction of the atom (or electron) with electromagnetic radiation, which it treats as a flux of quanta - photons.

One of the outstanding successes of Dirac's radiation theory was the derivation of the Einstein A-coefficient for spontaneous emission.

The representations of the discrete (quantum) nature of radiation and the related discreteness of all physical processes formed the basis of the interpretation of quantum mechanics and quantum electrodynamics; this is consistent with the initial ideas of M. Planck and A. Einstein. In particular, it is believed that any changes in the system occur in the form of the jump-like transitions in which the system goes from one allowed discrete state to another, while intermediate states do not exist.

In papers [1-6], a fundamentally different point of view of the nature of light, electrons and atoms has been suggested. According to this view, photons as the "particles of light" do not exist, and light is a continuous classical electromagnetic field fully described by Maxwell equations [1-3]. The electron as a particle also does not exist; in its place there is a continuous classical wave field - the electron wave having continuously distributed in space (i) its electric charge, (ii) its internal angular momentum, and (iii) its internal magnetic moment [4]. This field is completely described by the field equations - the Dirac equation, which plays for the electron wave the same role that the Maxwell equations play for the electromagnetic field. From this point of view, the electron field should be perceived as well as we perceive the classical electromagnetic field, with the only difference that the latter does not have its own electric charge, internal angular momentum and internal magnetic moment. Therein lies the difficulty of perception of an electron field because we are accustomed to the idea that electric charge, angular momentum and magnetic moments are the attributes of particles, and objects that possess them are immediately associated with particles.

From the point of view of the classical-field representations of the electron wave, the atom is an open volume resonator in which an electrically charged electron wave is held by the electrostatic field of the atomic nucleus [5].

As shown in [5], the external electrostatic field plays the role of a "dielectric medium" for the electron wave; "permittivity" is associated with the potential of the electrostatic field. From this perspective, it is also possible to say that the electron wave is held in the atom because of its total internal reflection in a non-uniform "dielectric medium", whose role is played by the electrostatic field of the nucleus.

From a formal point of view, an atom as an open volume resonator in which an electron wave is locked is no different from cavity resonators designed to hold electromagnetic waves. In particular, the atom, as a volume resonator, has a set of eigenmodes and corresponding eigenfrequencies [5]. The eigenfrequencies $\omega_n$ are the eigenvalues of the linear wave equation describing the electron wave (depending on the details of the process, it can be the Dirac equation, the Pauli equation, the Klein-Gordon equation or the Schrödinger equation), while the eigenmodes are described by the eigenfunctions $u_n(\mathbf{r})$ of this equation. Any state of the atom can be represented as a superposition of the eigenmodes

$$\psi = \sum_n c_n u_n(\mathbf{r}) \exp(-i\omega_n t) \tag{1}$$

where the eigenfunctions $u_n(\mathbf{r})$ are orthonormal:

$$\int u_n u_k^* dV = \delta_{nk} \tag{2}$$

while the coefficients $c_n$ describe the distribution of the electric charge of the electron wave between the eigenmodes in the atom [5].

In particular, for a hydrogen atom

$$\sum_n |c_n|^2 = 1 \tag{3}$$

while the value $q_n = -e|c_n|^2$ is the electric charge of the electron wave contained in eigenmode $n$. If the entire electric charge of the electron wave is contained in a single eigenmode $k$, that is $|c_k|^2 = 1$ and $|c_n|^2 = 0$ for $n \neq k$ (i.e., only one eigenmode $k$ is excited), then this state of the hydrogen atom is called a pure state; all other states of the hydrogen atom, when several eigenmodes are excited simultaneously, are called the mixed states [5].

The energy of the atom (of the electron wave) takes continuous values, although its eigenfrequencies $\omega_n$ have a discrete spectrum. This is similar to the notion that the energy of electromagnetic waves in a conventional cavity resonator can take continuous values, while its eigenfrequencies form a discrete spectrum.

I will discuss only the hydrogen atom, because at present, a self-consistent classical field theory has been developed only for it [5]. However, as will be shown in the following papers of this series, all of the results and conclusions obtained are applicable to other more complex atoms. According to classical electrodynamics, an atom in a pure state does not radiate electromagnetic waves, [5] and, therefore, can remain in this state indefinitely. That is, the pure states of the atom are stationary. In contrast, an atom in a mixed state emits electromagnetic waves continuously, in full compliance with classical electrodynamics [5]. In particular, if two eigenmodes $n$ and $k$ of an atom, having the eigenfrequencies $\omega_n$ and $\omega_k < \omega_n$, were excited simultaneously by any means, then the atom would *continuously* spontaneously emit electromagnetic waves at a frequency $\omega_{nk} = \omega_n - \omega_k$ [5]. This spontaneous emission will occur over time (i.e., it is not a jump-like transition as it is in quantum mechanics) and will be accompanied by a *continuous* change in the energy and the state of the atom: the electric charge of the electron wave in the process of spontaneous emission continuously flows from an eigenmode with a higher frequency $\omega_n$ into an eigenmode with a lower frequency $\omega_k$ (for brevity: from a higher mode to a lower mode) [5]. Thus, a mixed state of an atom is nonstationary, and because of spontaneous emission, the atom eventually will go into a stationary (pure) state corresponding to the lowest of the initially excited eigenmodes [5]. This implies that the lowest pure state of an atom, i.e., its ground state, is stable, while all other pure states, although they are stationary, are unstable.

If more than two eigenmodes of an atom are simultaneously excited, then the emission of the atom will occur *simultaneously at all frequencies* $\omega_{nk} = \omega_n - \omega_k > 0$, where $n$ and $k$ are the indices of all excited modes. This explains the discrete spectrum of the spontaneous emissions of atoms, which is not associated with jump-like transitions, but is a fairly trivial result of classical electrodynamics when the electrically charged electron wave is described by a wave equation, for example, the Schrödinger equation [5].

However, as we know, it is not possible to describe the spontaneous "transitions" themselves using a linear wave equation, for example, a linear Schrödinger equation because a linear wave equation does not describe the changes in the states of an atom in spontaneous emission. To explain spontaneous transitions in quantum theory, quantum mechanics is extended to quantum electrodynamics, in which not only the states of an atom are quantized but also the radiation itself.

In the paper [5], it is shown that this can be performed at the expense of the natural from the standpoint of a classical electrodynamics modification of the Schrödinger equation. So the equation obtained in [5]

$$i\hbar \frac{\partial \psi}{\partial t} = -\frac{1}{2m_e}\Delta\psi - \frac{e^2}{r}\psi - \frac{2e^2}{3c^3}\psi\mathbf{r}\frac{\partial^3}{\partial t^3}\int \mathbf{r}|\psi|^2 d\mathbf{r} \qquad (4)$$

allows describing not only stationary states of the hydrogen atom, which the conventional linear Schrödinger equation predicts but also the continuous changes in the states of an atom during spontaneous emission. The last term in the right-hand side of equation (4) is responsible for this, and it has a simple physical meaning from the standpoint of classical electrodynamics: this term describes the inverse action of the self-radiation field on the electron wave [5]. With the addition of the last term, the equation (4) becomes nonlinear.

This view allows (in a natural way within the framework of classical field theory) describing light-atom interactions with damping because of spontaneous emission and giving a fully classical explanation of the laws of the photoelectric effect [6].

Note that the interpretation proposed in [4,5] of the wave function and an electron wave described by it as a real physical field is close to the initial Schrödinger interpretation, which he abandoned under the influence of critics supporting the Copenhagen probabilistic interpretation. One of the arguments of opponents of the Schrödinger interpretation was that it will inevitably come into conflict with the Planck theory of thermal radiation requiring a discreteness of energy. In this paper, I will show that this is not so: the thermal radiation can be described as a purely classical process of interaction of a continuous electron wave in an atom with a continuous radiation field, being based on equation (4) and its generalization.

In this paper, I will give a detailed derivation of Planck's law for the equilibrium thermal radiation, considering the *continuous* process of interaction of a hydrogen atom with its thermal environment (thermostat) and with the radiation field generated by other atoms.

## 2 Atom in an Isotropic Light Field

2.1 Equations with damping due to spontaneous emission

Let us consider first an isolated atom embedded in an external radiation field created by other sources, and perhaps by it itself. In the paper [6], the case was considered in which an atom interacts with a plane monochromatic electromagnetic wave, and the optical equation with damping from spontaneous emission was rigorously derived. In present paper, we assume that the radiation field is not monochromatic and is a superposition of waves with a broad frequency spectrum having different polarizations and propagating in different directions.

The hydrogen atom in an electromagnetic wave field is described by the nonlinear Schrödinger equation [5,6]

$$i\hbar \frac{\partial \psi}{\partial t} = -\frac{1}{2m_e} \Delta \psi - e\varphi\psi + \psi e\mathbf{r}\mathbf{E}(\mathbf{r},t) + \frac{2e}{3c^3}\psi \mathbf{r}\ddot{\mathbf{d}} \qquad (5)$$

where

$$\mathbf{d} = -e \int \mathbf{r}|\psi|^2 d\mathbf{r} \qquad (6)$$

is the electric dipole moment of the electron wave in an atom; $\mathbf{E}(\mathbf{r},t)$ is the electric field of an electromagnetic wave, which in the case under consideration is the result of interference of a plurality of the waves created by different sources (atoms).

In this paper, it is assumed that the wavelengths $\lambda$ of the electromagnetic radiation are substantially larger than the characteristic size of the electron field in the atom:

$$\lambda \gg \int r|\psi|^2 d\mathbf{r} \qquad (7)$$

Therefore, the coordinates $\mathbf{r}$ in the vector $\mathbf{E}(\mathbf{r},t)$ in solving equation (5) are assumed to be constant and are treated as parameters, and for this reason are not indicated further.

If the atom is in a mixed state, its wave function has the form (1). Substituting (1) into equation (5), one obtains, taking into account expression (2), the system of equations for the amplitudes of the excitation of eigenmodes [5]

$$i\hbar \frac{dc_n}{dt} = -\sum_k c_k (\mathbf{d}_{nk}\mathbf{E}) \exp(-i\omega_{kn}t) - \frac{2}{3c^3}\sum_k c_k (\mathbf{d}_{nk}\ddot{\mathbf{d}}) \exp(-i\omega_{kn}t) \qquad (8)$$

where $\omega_{nk} = \omega_n - \omega_k$;

$$\ddot{\mathbf{d}} = i\sum_n \sum_k \omega_{nk}^3 c_n c_k^* \mathbf{d}_{nk}^* \exp(-i\omega_{nk}t) \qquad (9)$$

$$\mathbf{d}_{nk} = \mathbf{d}_{kn}^* = -e \int \mathbf{r} u_n^*(\mathbf{r}) u_k(\mathbf{r}) d\mathbf{r} \qquad (10)$$

It is assumed that the amplitudes $c_n$ change slowly compared with the rapidly oscillating factors $\exp(\pm i\omega_{nk}t)$, i.e.,

$$|\dot{c}_n| \ll \omega_{nk}|c_n| \qquad (11)$$

Justification of this assumption was given in [6].

The solution of equations (8), (9) allows describing the behaviour of atoms in any given radiation field $\mathbf{E}(t)$.

As an example, let us consider the so-called "two-level atom", i.e., the case in which only two eigenmodes $k$ and $n$ of the atom are excited simultaneously. For concreteness, assume that $\omega_n > \omega_k$, i.e., $\omega_{nk} > 0$.

In this case, equations (8) and (9) take the form

$$i\hbar \frac{dc_n}{dt} = -[c_k(\mathbf{d}_{nk}\mathbf{E})\exp(i\omega_{nk}t) + c_n(\mathbf{d}_{nn}\mathbf{E})] - \frac{2}{3c^3}[c_k(\mathbf{d}_{nk}\ddot{\mathbf{d}})\exp(i\omega_{nk}t) + c_n(\mathbf{d}_{nn}\ddot{\mathbf{d}})] \quad (12)$$

$$i\hbar \frac{dc_k}{dt} = -[c_n(\mathbf{d}_{nk}^*\mathbf{E})\exp(-i\omega_{nk}t) + c_k(\mathbf{d}_{kk}\mathbf{E})] - \frac{2}{3c^3}[c_n(\mathbf{d}_{nk}^*\ddot{\mathbf{d}})\exp(-i\omega_{nk}t) + c_k(\mathbf{d}_{kk}\ddot{\mathbf{d}})]$$

$$(13)$$

$$\ddot{\mathbf{d}} = i\omega_{nk}^3[\mathbf{d}_{nk}^* c_n c_k^* \exp(-i\omega_{nk}t) - \mathbf{d}_{nk} c_k c_n^* \exp(i\omega_{nk}t)] \qquad (14)$$

In what follows we assume that the field $\mathbf{E}(t)$ corresponds to electromagnetic waves, and thus it is oscillating. Then, equations (12) and (13) contain pure oscillating terms that cannot be compensated in any way. However, the amplitudes $c_n$ and $c_k$ will contain both the rapidly oscillating components that vary with frequency $\omega_{nk}$ or with the characteristic frequencies of the field $\mathbf{E}(t)$ and components varying slowly compared to them. In this paper we are interested in the slowly varying components of the amplitudes $c_n$ and $c_k$. Therefore, averaging equations (12) and (13) over the fast oscillations, one obtains

$$i\hbar \frac{dc_n}{dt} = -c_k \overline{(\mathbf{d}_{nk}\mathbf{E})\exp(i\omega_{nk}t)} - ic_n|c_k|^2 \frac{2\omega_{nk}^3}{3c^3}|\mathbf{d}_{nk}|^2 \quad (15)$$

$$i\hbar \frac{dc_k}{dt} = -c_n \overline{(\mathbf{d}_{nk}^*\mathbf{E})\exp(-i\omega_{nk}t)} + ic_k|c_n|^2 \frac{2\omega_{nk}^3}{3c^3}|\mathbf{d}_{nk}|^2 \quad (16)$$

where the bar denotes averaging over the fast oscillations.

Let us consider the spectral representation of the electromagnetic radiation field

$$\mathbf{E}(t) = \int_{-\infty}^{\infty} \mathbf{E}_\omega(\omega)\exp(-i\omega t)\,d\omega \quad (17)$$

where

$$\mathbf{E}_\omega(-\omega) = \mathbf{E}_\omega^*(\omega) \quad (18)$$

Next, one can write

$$\mathbf{E}(t) = \int_0^\infty \mathbf{E}_\omega(\omega)\exp(-i\omega t)\,d\omega + \int_0^\infty \mathbf{E}_\omega^*(\omega)\exp(i\omega t)\,d\omega \quad (19)$$

Substituting (19) into equations (15) and (16), one obtains

$$i\hbar \frac{dc_n}{dt} = -c_k \int_0^\infty (\mathbf{d}_{nk}\mathbf{E}_\omega)\exp[-i(\omega-\omega_{nk})t]\,d\omega - ic_n|c_k|^2 \frac{2\omega_{nk}^3}{3c^3}|\mathbf{d}_{nk}|^2 \quad (20)$$

$$i\hbar \frac{dc_k}{dt} = -c_n \int_0^\infty (\mathbf{d}_{nk}^*\mathbf{E}_\omega^*)\exp[i(\omega-\omega_{nk})t]\,d\omega + ic_k|c_n|^2 \frac{2\omega_{nk}^3}{3c^3}|\mathbf{d}_{nk}|^2 \quad (21)$$

where it is assumed that

$$|\omega - \omega_{nk}| \ll \omega_{nk} \quad (22)$$

Therefore, when averaging over the fast oscillations, the factors $\exp[\pm i(\omega - \omega_{nk})t]$ are considered to be constant. Condition (22) will be proved later.

Let us introduce the notation

$$\rho_{nn} = |c_n|^2, \rho_{kk} = |c_k|^2, \rho_{nk} = c_n c_k^*, \rho_{kn} = c_k c_n^* \quad (23)$$

Obviously,

$$\rho_{nn} + \rho_{kk} = 1 \quad (24)$$

$$\rho_{nk} = \rho_{kn}^* \quad (25)$$

Condition (24) expresses the law of conservation of charge for the electron wave [4,5], and means that in the process under consideration, the electric charge of the electron wave is distributed between only the modes $n$ and $k$.

Using equations (20) and (21), one obtains

$$\frac{d\rho_{nn}}{dt} = -\frac{d\rho_{kk}}{dt} = i\frac{1}{\hbar}\{\rho_{kn}\int_0^\infty (\mathbf{d}_{nk}\mathbf{E}_\omega)\exp[-i(\omega-\omega_{nk})t]\,d\omega - \rho_{nk}\int_0^\infty (\mathbf{d}_{nk}^*\mathbf{E}_\omega^*)\exp[i(\omega-\omega_{nk})t]\,d\omega\} - 2\gamma_{nk}\rho_{nn}\rho_{kk} \quad (26)$$

$$\frac{d\rho_{nk}}{dt} = \frac{d\rho_{kn}^*}{dt} = (\rho_{kk}-\rho_{nn})\{i\frac{1}{\hbar}\int_0^\infty (\mathbf{d}_{nk}\mathbf{E}_\omega)\exp[-i(\omega-\omega_{nk})t]\,d\omega - \gamma_{nk}\rho_{nk}\} \quad (27)$$

where

$$\gamma_{nk} = \frac{2\omega_{nk}^3}{3\hbar c^3}|\mathbf{d}_{nk}|^2 \quad (28)$$

Equations (26) and (27) are a generalisation of the equations obtained in [6] for the case when the electromagnetic wave field has a continuous spectrum described by the spectral function $\mathbf{E}_\omega(\omega)$.

Let us assume that the functions $\rho_{nn}$ and $\rho_{kk}$ are changed slowly compared with the oscillating functions $\rho_{nk}$ and $\rho_{kn}$, so that while functions $\rho_{nk}$ and $\rho_{kn}$ are changing, one can ignore any changes in functions $\rho_{nn}$ and $\rho_{kk}$. In this case, equation (27) has a solution

$$\rho_{nk} = \int_0^\infty a_\omega \exp[-i(\omega-\omega_{nk})t]\,d\omega \quad (29)$$

where $a_\omega$ is a slowly varying function of time.

Substituting (29) into (27) one obtains

$$a_\omega = i\frac{(\rho_{kk}-\rho_{nn})}{-i(\omega-\omega_{nk})+(\rho_{kk}-\rho_{nn})\gamma_{nk}}\frac{(\mathbf{d}_{nk}\mathbf{E}_\omega)}{\hbar} \quad (30)$$

Substituting (29) and (30) into (26) and taking into account (25), one obtains

$$\frac{d\rho_{nn}}{dt} = -\frac{d\rho_{kk}}{dt} = \frac{1}{\hbar}\{\int_0^\infty\int_0^\infty \frac{(\rho_{kk}-\rho_{nn})}{i(\omega'-\omega_{nk})+(\rho_{kk}-\rho_{nn})\gamma_{nk}}\frac{(\mathbf{d}_{nk}^*\mathbf{E}_{\omega'}^*)(\mathbf{d}_{nk}\mathbf{E}_\omega)}{\hbar}\exp[-i(\omega-\omega')t]\,d\omega'\,d\omega +$$

$$\int_0^\infty\int_0^\infty \frac{(\rho_{kk}-\rho_{nn})}{-i(\omega'-\omega_{nk})+(\rho_{kk}-\rho_{nn})\gamma_{nk}}\frac{(\mathbf{d}_{nk}\mathbf{E}_{\omega'})(\mathbf{d}_{nk}^*\mathbf{E}_\omega^*)}{\hbar}\exp[i(\omega-\omega')t]\,d\omega'\,d\omega \quad (31)$$

Let us average this equation over the time interval $2t_0$, which is assumed to be considerably less than the characteristic time of change of the slowly changing functions $\rho_{nn}$ and $\rho_{kk}$, but much longer than the characteristic time of change of the rapidly oscillating function $\rho_{nk}$. Taking into account that as $t_0 \to \infty$

$$\overline{\exp[\pm i(\omega-\omega')t]} = \frac{1}{2t_0}\int_{-t_0}^{t_0}\exp[\pm i(\omega-\omega')t]\,dt = \frac{\pi}{t_0}\delta(\omega-\omega') \quad (32)$$

one obtains as a result

$$\frac{d\rho_{nn}}{dt} = -\frac{d\rho_{kk}}{dt} = \frac{2\pi}{\hbar^2}\frac{1}{t_0}\int_0^\infty \frac{(\rho_{kk}-\rho_{nn})^2\gamma_{nk}}{(\omega-\omega_{nk})^2+(\rho_{kk}-\rho_{nn})^2\gamma_{nk}^2}(\mathbf{d}_{nk}^*\mathbf{E}_\omega^*)(\mathbf{d}_{nk}\mathbf{E}_\omega)\,d\omega - 2\gamma_{nk}\rho_{nn}\rho_{kk} \quad (33)$$

Let us calculate the mean energy density of the radiation field

$$W = \frac{1}{4\pi}\overline{\mathbf{E}^2} \quad (34)$$

We assume that the radiation field exists only during the time interval $[-t_0, t_0]$, and then pass to the limit $t_0 \to \infty$.

Then, taking (19) and (32) into account, one obtains

$$W = \frac{1}{2t_0} \int_0^\infty \overline{|\mathbf{E}_\omega(\omega)|^2} d\omega \quad (35)$$

This expression can be written in the form

$$W = \int_0^\infty U_\omega(\omega) d\omega \quad (36)$$

where

$$U_\omega(\omega) = \frac{1}{2t_0} \overline{|\mathbf{E}_\omega(\omega)|^2} \quad (37)$$

is the spectral energy density: $U_\omega(\omega)d\omega$ is the energy of radiation field having a frequency in the range $[\omega, \omega + d\omega]$, per unit volume.

For an isotropic radiation field,

$$\overline{E_{\omega i}(\omega) E^*_{\omega j}(\omega)} = \frac{1}{3} \overline{|\mathbf{E}_\omega(\omega)|^2} \delta_{ij} \quad (38)$$

where $i, j = 1,2,3$ are the vector indices.

Taking into account random orientations of the electric dipole moments $\mathbf{d}_{nk}$ with respect to the vectors $\mathbf{E}_\omega$ and their statistical independence, one can write

$$\overline{\langle (\mathbf{d}^*_{nk} \mathbf{E}^*_\omega)(\mathbf{d}_{nk} \mathbf{E}_\omega) \rangle} = \langle d_{nk,i} d^*_{nk,j} \rangle \overline{E_{\omega i} E^*_{\omega j}} \quad (39)$$

where $\langle ... \rangle$ denotes averaging over all mutual orientations of the vectors $\mathbf{d}_{nk}$ and $\mathbf{E}_\omega$; $i$ and $j$ are the vector indices; assume summation over repeated vector indices.

Because all orientations of the vector $\mathbf{d}_{nk}$ are equally probable, one obtains

$$\langle d_{nk,i} d^*_{nk,j} \rangle = \frac{1}{3} |\mathbf{d}_{nk}|^2 \delta_{ij} \quad (40)$$

Then, taking (38) into account, one can write expression (39) in the form

$$\overline{\langle (\mathbf{d}^*_{nk} \mathbf{E}^*_\omega)(\mathbf{d}_{nk} \mathbf{E}_\omega) \rangle} = \frac{1}{3} |\mathbf{d}_{nk}|^2 \overline{|\mathbf{E}_\omega|^2} \quad (41)$$

Averaging equation (33) over all possible mutual orientations of the vectors $\mathbf{d}_{nk}$ and $\mathbf{E}_\omega$, taking expressions (37) and (41) into account, one obtains

$$\frac{d\rho_{nn}}{dt} = -\frac{d\rho_{kk}}{dt} = \frac{4\pi}{3\hbar^2} |\mathbf{d}_{nk}|^2 \int_0^\infty \frac{(\rho_{kk} - \rho_{nn})^2 \gamma_{nk}}{(\omega - \omega_{nk})^2 + (\rho_{kk} - \rho_{nn})^2 \gamma_{nk}^2} U_\omega(\omega) d\omega - 2\gamma_{nk} \rho_{nn} \rho_{kk} \quad (42)$$

This is the most general equation describing the behaviour of the two-level atom in an isotropic light field. Knowing the spectral energy density of the radiation field $U_\omega(\omega)$ in which an atom finds itself, one can calculate the parameters $\rho_{nn}$ and $\rho_{kk}$ using equation (42).

In particular, an atom will be in equilibrium with the radiation field when $\frac{d\rho_{nn}}{dt} = \frac{d\rho_{kk}}{dt} = 0$, i.e., when

$$\frac{\pi c^3}{\hbar \omega_{nk}^3} \int_0^\infty \frac{(1 - 2\rho_{nn})^2 \gamma_{nk}}{(\omega - \omega_{nk})^2 + (1 - 2\rho_{nn})^2 \gamma_{nk}^2} U_\omega(\omega) d\omega = \rho_{nn}(1 - \rho_{nn}) \quad (43)$$

Here, expressions (24) and (28) are taken into account. Equation (43) allows finding the distribution of the electric charge of the electron wave between the excited modes $n$ and $k$ (i.e., the parameters $\rho_{nn}$ and $\rho_{kk}$) for the case when the atom is in equilibrium with radiation characterised by the spectral energy density $U_\omega(\omega)$.

It is noteworthy that the equilibrium condition (43) does not contain an electric dipole moment $|\mathbf{d}_{nk}|^2$, and none of the parameters that characterise the excited states of the atom other than frequency $\omega_{nk}$. This means that the equilibrium is universal and is determined only by the frequency $\omega_{nk}$.

## 2.2 Continuous spectrum of the radiation field

Let us assume that there are many sources of radiation (the atoms) with different frequencies of spontaneous emission, so that together they form a continuous spectrum of the radiation field. Next, equation (42) describes the excitation of the two-level atom, which has a frequency of spontaneous emission $\omega_{nk}$, in this light field.

Taking into account the assessment [6]

$$\gamma_{nk}/\omega_{nk} \sim \alpha^3 \ll 1 \tag{44}$$

where $\alpha = \frac{e^2}{\hbar c}$ is the fine-structure constant, one concludes that the function $\frac{(\rho_{kk}-\rho_{nn})^2 \gamma_{nk}}{(\omega-\omega_{nk})^2 + (\rho_{kk}-\rho_{nn})^2 \gamma_{nk}^2}$ has a bell-shaped form in the vicinity of the frequency $\omega_{nk}$ with bandwidth $\Delta\omega \sim \gamma_{nk} \ll \omega_{nk}$. This proves condition (22).

By the term continuous spectrum of the radiation field, we will mean a spectrum whose change in spectral energy density $U_\omega(\omega)$ in the band $\Delta\omega \sim \gamma_{nk}$ in the vicinity of frequency $\omega_{nk}$ is small compared with the value of $U_\omega(\omega)$, i.e.,

$$\gamma_{nk} \left|\frac{dU_\omega(\omega_{nk})}{d\omega}\right| \ll U_\omega(\omega_{nk}) \tag{45}$$

In this case, taking into account the expressions (44) and (45), one obtains

$$\int_0^\infty \frac{(\rho_{kk}-\rho_{nn})^2 \gamma_{nk}}{(\omega-\omega_{nk})^2 + (\rho_{kk}-\rho_{nn})^2 \gamma_{nk}^2} U_\omega(\omega) d\omega \approx U_\omega(\omega_{nk}) \int_0^\infty \frac{(\rho_{kk}-\rho_{nn})^2 \gamma_{nk}}{(\omega-\omega_{nk})^2 + (\rho_{kk}-\rho_{nn})^2 \gamma_{nk}^2} d\omega \approx$$

$$U_\omega(\omega_{nk}) \int_{-\infty}^\infty \frac{(\rho_{kk}-\rho_{nn})^2 \gamma_{nk}}{w^2 + (\rho_{kk}-\rho_{nn})^2 \gamma_{nk}^2} dw = \pi U_\omega(\omega_{nk})(\rho_{kk}-\rho_{nn}) \tag{46}$$

Substituting expression (46) into equation (42), one obtains

$$\frac{d\rho_{nn}}{dt} = -\frac{d\rho_{kk}}{dt} = \frac{4\pi^2 |\mathbf{d}_{nk}|^2}{3\hbar^2} U_\omega(\omega_{nk})(\rho_{kk}-\rho_{nn}) - 2\gamma_{nk}\rho_{nn}\rho_{kk} \tag{47}$$

Taking into account expression (24), equation (47) can be rewritten in the form

$$\frac{d\rho_{nn}}{dt} = \frac{4\pi^2 |\mathbf{d}_{nk}|^2}{3\hbar^2} U_\omega(\omega_{nk})(1 - 2\rho_{nn}) - 2\gamma_{nk}\rho_{nn}(1-\rho_{nn}) \tag{48}$$

The solution of this equation with the initial condition

$$\rho_{nn}(t=0) = \rho_{nn}(0) \tag{49}$$

has the form

$$\rho_{nn} = \frac{\rho_{nn}(0)\{p - q\exp[2\gamma_{nk}(p-q)t]\} - pq\{1 - \exp[2\gamma_{nk}(p-q)t]\}}{\rho_{nn}(0)\{1 - \exp[2\gamma_{nk}(p-q)t]\} + p\exp[2\gamma_{nk}(p-q)t] - q} \tag{50}$$

where

$$p = \frac{1}{2}(1+2a) + \frac{1}{2}\sqrt{1+4a^2}, \qquad q = \frac{1}{2}(1+2a) - \frac{1}{2}\sqrt{1+4a^2} \tag{51}$$

are the roots of the equation

$$\rho_{nn}^2 - (1+2a)\rho_{nn} + a = 0 \tag{52}$$

$$a = \frac{2\pi^2|\mathbf{d}_{nk}|^2}{3\hbar^2 \gamma_{nk}} U_\omega(\omega_{nk}) \tag{53}$$

Using (28), one obtains

$$a = \frac{\pi^2 c^3}{\hbar \omega_{nk}^3} U_\omega(\omega_{nk}) \tag{54}$$

Equation (48) has two stationary points, $\rho_{nn} = q$ and $\rho_{nn} = p$. It is easy to see that the first is stable, while the second does not make sense because $p > 1$; i.e., it does not correspond to any state of the neutral hydrogen atom.

From expression (50) it follows that at $t \to \infty$

$$\rho_{nn}(\infty) = q \tag{55}$$

In this state, the atom will be in equilibrium with the radiation field.

Thus, in equilibrium state, the atom will have the following distribution of the electric charge of the electron wave between the excited eigenmodes $n$ and $k$

$$\rho_{nn} = \frac{1}{2}(1+2a) - \frac{1}{2}\sqrt{1+4a^2} \tag{56}$$

$$\rho_{kk} = \frac{1}{2}(1-2a) + \frac{1}{2}\sqrt{1+4a^2} \tag{57}$$

These expressions can be obtained directly by solving the algebraic equation (52), which is obtained by equating the right-hand side of equation (48) to zero, which corresponds to the equilibrium state of the atom.

Note that the initial condition (49) can be created by different types of exposure, for example, by the thermal collisions of the atom under study with other atoms, by laser radiation, etc. The excitation of the atoms because of "thermal collisions" with other atoms will be discussed below.

2.3 The energy condition of the equilibrium of the atom and the radiation field

By expressions (56) and (57), both eigenmodes $n$ and $k$ of the two-level atom will be excited simultaneously in an equilibrium state. This means that the atom, being in equilibrium with the radiation field, will be in a mixed excited state, and according to [5] will continuously emit electromagnetic waves in full compliance with classical electrodynamics. On the other hand, again in full compliance with classical electrodynamics, the radiation field will do work on the distributed electric currents of the electron wave in an atom, which will lead to the excitation of the atom. Obviously, in equilibrium conditions of the atom in a radiation field, the energy balance condition must be satisfied: the energy emitted by the atom in the form of electromagnetic waves must be equal to the energy absorbed by the atom from the light field:

$$I = W_E \tag{58}$$

where $I$ is the intensity of the emission of an atom, and $W_E$ is the mean energy absorbed by the electron wave from of the radiation field per unit time.

According to classical electrodynamics [7]

$$I = \frac{2}{3c^3} \overline{\ddot{\mathbf{d}}^2} \tag{59}$$

$$W_E = \int \overline{\mathbf{j}\mathbf{E}} dV \tag{60}$$

Condition (7) can be written as

$$W_E = \overline{\mathbf{E} \int \mathbf{j} dV} \tag{61}$$

From the continuity equation for the electric charge of the electron wave continuously distributed in space, it is easy to obtain

$$\int \mathbf{j} dt = \dot{\mathbf{d}} \tag{62}$$

Next, expression (61) can be rewritten in the form

$$W_E = \overline{\mathbf{E}\dot{\mathbf{d}}} \tag{61}$$

Using condition (11), one obtains

$$\dot{\mathbf{d}} = -i \sum_n \sum_k \omega_{nk} c_n c_k^* \mathbf{d}_{nk}^* \exp(-i\omega_{nk} t) \tag{62}$$

$$\ddot{\mathbf{d}} = - \sum_n \sum_k \omega_{nk}^2 c_n c_k^* \mathbf{d}_{nk}^* \exp(-i\omega_{nk} t) \tag{63}$$

Then

$$I = \frac{2}{3c^3} \sum_n \sum_k \omega_{nk}^4 \rho_{nn} \rho_{kk} |\mathbf{d}_{nk}|^2 \tag{64}$$

and

$$W_E = -i \sum_n \sum_k \omega_{nk} \mathbf{d}_{nk}^* \cdot \overline{\rho_{nk} \mathbf{E} \exp(-i\omega_{nk} t)} \tag{65}$$

Substituting expressions (64) and (65) into the energy balance equation (58), one obtains the general equilibrium condition of the atom with the radiation field.

In particular, for a two-level atom, using (10) and (23) one obtains

$$I = \frac{4\omega_{nk}^4 |\mathbf{d}_{nk}|^2}{3c^3} \rho_{nn} \rho_{kk} \tag{66}$$

$$W_E = -i\omega_{nk} \mathbf{d}_{nk}^* \cdot \overline{\rho_{nk} \mathbf{E} \exp(-i\omega_{nk} t)} + i\omega_{nk} \mathbf{d}_{nk} \cdot \overline{\rho_{nk}^* \mathbf{E} \exp(i\omega_{nk} t)} \tag{67}$$

With expression (28), (66) can be written in the form

$$I = 2\gamma_{nk}\hbar\omega_{nk}\rho_{nn}\rho_{kk} \tag{68}$$

Using expressions (19) and (23) and solution (29) and (30), one obtains

$$\overline{\rho_{nk}\mathbf{E}\exp(-i\omega_{nk}t)} = \int_0^\infty\int_0^\infty \mathbf{E}_\omega^*(\omega)a_{\omega'}\overline{\exp[i(\omega-\omega')t]}\,d\omega'\,d\omega = \frac{\pi}{t_0}\int_0^\infty \mathbf{E}_\omega^*(\omega)a_\omega d\omega \tag{69}$$

Substituting this expression into (67), one obtains

$$W_E = \omega_{nk}\frac{2\pi}{\hbar t_0}\int_0^\infty \frac{(\rho_{kk}-\rho_{nn})^2\gamma_{nk}}{(\omega-\omega_{nk})^2+(\rho_{kk}-\rho_{nn})^2\gamma_{nk}^2}(\mathbf{d}_{nk}^*\mathbf{E}_\omega^*)(\mathbf{d}_{nk}\mathbf{E}_\omega)d\omega \tag{70}$$

Averaging expression (70) over all possible mutual orientations of the vectors $\mathbf{d}_{nk}$ and $\mathbf{E}_\omega$, and using expressions (37) and (41), one obtains

$$W_E = \omega_{nk}\frac{4\pi|\mathbf{d}_{nk}|^2}{3\hbar}\int_0^\infty \frac{(\rho_{kk}-\rho_{nn})^2\gamma_{nk}}{(\omega-\omega_{nk})^2+(\rho_{kk}-\rho_{nn})^2\gamma_{nk}^2}U_\omega(\omega)d\omega \tag{71}$$

For the continuous spectrum of the radiation field, using (46), one obtains

$$W_E = \hbar\omega_{nk}\frac{4\pi^2|\mathbf{d}_{nk}|^2}{3\hbar^2}U_\omega(\omega_{nk})(\rho_{kk}-\rho_{nn}) \tag{72}$$

Substituting expressions (68) and (72) into the energy balance condition (58), one obtains

$$2\gamma_{nk}\rho_{nn}\rho_{kk} = \frac{4\pi^2|\mathbf{d}_{nk}|^2}{3\hbar^2}U_\omega(\omega_{nk})(\rho_{kk}-\rho_{nn}) \tag{73}$$

Equation (73) coincides exactly with the condition of equilibrium of the atom with the radiation field, which was obtained from the equation (47) at $\frac{d\rho_{nn}}{dt} = \frac{d\rho_{kk}}{dt} = 0$. Thus, the energy equilibrium condition of the atom with the radiation field (73) is also a consequence of equation (47). One can conclude therefore that equation (47), obtained from the nonlinear Schrödinger equation (5), which takes spontaneous emission into account, contains all of the necessary energy conditions of the interaction of the atom with the radiation field.

If the spectral energy density of the radiation field is known, then the distribution of the electric charge of the electron wave between the excited modes of the atom in equilibrium can be determined with expressions (56) and (57).

Conversely, if the distribution of the electric charge of the electron wave between the excited eigenmodes of an atom in equilibrium with the radiation field is known *a priori*, then from expression (73) and (28) one can find the spectral energy density of radiation at which such a balance is possible:

$$U_\omega(\omega_{nk}) = \frac{\hbar\omega_{nk}^3}{\pi^2 c^3}\frac{\rho_{nn}\rho_{kk}}{\rho_{kk}-\rho_{nn}} \tag{74}$$

Note that in this case, the reason that the atom turned up in a mixed state characterised by the parameters $\rho_{nn}$ and $\rho_{kk}$ was not specified; therefore, expression (74) is valid for any reason causing the excitation of its eigenmodes. In the case of thermal excitation of the atom, it is necessary to use the thermostatistics of excitation of various eigenmodes of the electron wave in an atom.

## 3  Interaction of an Atom with a Thermal Reservoir

3.1  Schrödinger-Langevin equation for the hydrogen atom

Equation (5) cannot describe the thermal excitation of an atom from its thermal collisions with the atoms of a thermal reservoir.

To take the thermal effect on the atom into account, it is necessary to add to equation (5) a "collision term" that describes the additional external potential field $\varphi_T(\mathbf{r},t)$ created inside the given atom by other atoms in its thermal environment, or by the atoms of a thermal reservoir due to thermal collisions. In this case, the Schrödinger equation (5) takes the form

$$i\hbar \frac{\partial \psi}{\partial t} = -\frac{1}{2m}\Delta\psi - \frac{e^2}{r}\psi - e\varphi_T(\mathbf{r},t)\psi + \psi e\mathbf{r}\mathbf{E}(t) - \frac{2e^2}{3c^3}\psi\mathbf{r}\frac{\partial^3}{\partial t^3}\int \mathbf{r}|\psi|^2 d\mathbf{r} \qquad (75)$$

The potential $\varphi_T(\mathbf{r},t)$ plays a role of a random external "force", leading to excitation of the different eigenmodes of the atom under consideration from thermal collisions with other atoms. Because of these collisions, the transition of an atom into a mixed (excited) state occurs that is accompanied by the continuous redistribution of the electric charge of the electron wave between the different eigenmodes of the atom. Once in the mixed state, the atom emits electromagnetic waves in full compliance with classical electrodynamics [5]. It follows that *the natures of the thermal and the spontaneous emissions of an atom are the same: both are the classical electric dipole emission of the atom in a mixed state* [5]. During a spontaneous emission, the atom tends to return to the ground state, in which only one lowest eigenmode is excited; however, the random thermal effects do not allow it to do this. If there is an external radiation field $\mathbf{E}(t)$, then it will lead to additional excitations of the different eigenmodes of the atom, which will also contribute to its transition to a mixed state. This is usually interpreted as a stimulated emission. All these processes are described by equation (6), which, in fact, is the quantum analogue of the Langevin equation.

Let us consider first the case in which the electromagnetic waves emitted by the atoms propagate to infinity. As a result, there is no radiation field around the atom. Such a situation is possible when the radiation created by the atoms is not retained in the part of space containing atoms emitting electromagnetic waves. In this case, the atoms are influenced only by their thermal collisions (thermal excitations). This process is described by the Schrödinger-Langevin equation (75), in which there is no external electromagnetic field:

$$i\hbar \frac{\partial \psi}{\partial t} = -\frac{1}{2m} \Delta \psi - e\varphi\psi - e\varphi_T(\mathbf{r},t)\psi - \frac{2e^2}{3c^3}\psi\mathbf{r}\frac{\partial^3}{\partial t^3}\int \mathbf{r}|\psi|^2 d\mathbf{r} \quad (76)$$

This equation takes into account the thermal excitation of an atom and its spontaneous emission. Substituting wave function (1) into equation (76) and using expression (2), one obtains the system of equations for the amplitudes of the excitations of the eigenmodes of the atom $c_n$. In particular, for a two-level atom, the system of equations has the form

$$i\hbar\frac{dc_n}{dt} =$$
$$c_k V_{nk}(t)\exp(i\omega_{nk}t) + c_n V_{nn}(t) - i\frac{2\omega_{nk}^3}{3c^3}c_n|c_k|^2|\mathbf{d}_{nk}|^2 -$$
$$i\frac{2\omega_{nk}^3}{3c^3}[c_k^* c_n^2(\mathbf{d}_{nn}\mathbf{d}_{nk}^*)\exp(-i\omega_{nk}t) + c_k|c_n|^2(\mathbf{d}_{nn}\mathbf{d}_{nk})\exp(i\omega_{nk}t)] +$$
$$i\frac{2\omega_{nk}^3}{3c^3}c_n^* c_k^2(\mathbf{d}_{nk}\mathbf{d}_{nk})\exp(2i\omega_{nk}t) \quad (77)$$

$$i\hbar\frac{dc_k}{dt} =$$
$$c_n V_{nk}^*(t)\exp(-i\omega_{nk}t) + c_k V_{kk}(t) - i\frac{2\omega_{nk}^3}{3c^3}c_k|c_n|^2|\mathbf{d}_{nk}|^2 -$$
$$i\frac{2\omega_{nk}^3}{3c^3}[c_n|c_k|^2(\mathbf{d}_{kk}\mathbf{d}_{nk}^*)\exp(-i\omega_{nk}t) + c_n^* c_k^2(\mathbf{d}_{kk}\mathbf{d}_{nk})\exp(i\omega_{nk}t)] -$$
$$i\frac{2\omega_{nk}^3}{3c^3}c_k^* c_n^2(\mathbf{d}_{nk}^*\mathbf{d}_{nk}^*)\exp(-2i\omega_{nk}t) \quad (78)$$

where

$$V_{nk}(t) = V_{kn}^*(t) = -e\int \varphi_T(\mathbf{r},t)u_n^*(\mathbf{r})u_k(\mathbf{r})d\mathbf{r} \quad (79)$$

The real-valued parameter $V_{nn}$ is the potential energy of the electron wave in a pure mode $n$ in the external field $\varphi_T(\mathbf{r},t)$.

The system of equations (77), (78) has rapidly oscillating terms that are not associated with the thermal interaction. These terms can be discarded as a result of averaging over the fast oscillations. As a result, with expression (28), one obtains

$$i\hbar\frac{dc_n}{dt} = c_k V_{nk}(t)\exp(i\omega_{nk}t) + c_n V_{nn}(t) - i\hbar\gamma_{nk}c_n|c_k|^2 \quad (80)$$

$$i\hbar\frac{dc_k}{dt} = c_n V_{nk}^*(t)\exp(-i\omega_{nk}t) + c_k V_{kk}(t) + i\hbar\gamma_{nk}c_k|c_n|^2 \quad (81)$$

Note that in the general case, when all eigenmodes of the atom can be excited, the equations (80), (81) take the form

$$i\hbar\frac{dc_n}{dt} = c_n V_{nn}(t) + \sum_{k\neq n} c_k V_{nk}(t)\exp(i\omega_{nk}t) - i\hbar c_n \sum_{k\neq n}\gamma_{nk}|c_k|^2 \quad (82)$$

where $n$ runs over all possible integer values.

Using the parameters (23), (25), one can write the equations (82) in the form

$$\frac{d\rho_{nn}}{dt} = -i\frac{1}{\hbar}\sum_{k\neq n}[\rho_{nk}^*V_{nk}(t)\exp(i\omega_{nk}t) - \rho_{nk}V_{nk}^*(t)\exp(-i\omega_{nk}t)] - 2\rho_{nn}\sum_{k\neq n}\gamma_{nk}\rho_{kk} \quad (83)$$

while for $n \neq k$

$$\frac{d\rho_{nk}}{dt} = -i\frac{1}{\hbar}\rho_{nk}[V_{nn}(t) - V_{kk}(t)] - i\frac{1}{\hbar}(\rho_{kk} - \rho_{nn})V_{nk}(t)\exp(i\omega_{nk}t) - (\rho_{kk} - \rho_{nn})\gamma_{nk}\rho_{nk} -$$

$$i\frac{1}{\hbar}\sum_{s\neq n,k}[\rho_{sk}V_{ns}(t)\exp(i\omega_{ns}t) - \rho_{ns}V_{ks}^*(t)\exp(-i\omega_{ks}t)] + \rho_{nk}\sum_{s\neq n,k}(\gamma_{ks} - \gamma_{ns})\rho_{ss} \quad (84)$$

At the same time, in accordance with (3)

$$\sum_n \rho_{nn} = 1 \quad (85)$$

For the two-level atom, taking note of expressions (24) and (25)

$$\frac{d\rho_{nn}}{dt} = -i\frac{1}{\hbar}[\rho_{nk}^*V_{nk}(t)\exp(i\omega_{nk}t) - \rho_{nk}V_{nk}^*(t)\exp(-i\omega_{nk}t)] - 2\rho_{nn}\gamma_{nk}\rho_{kk} \quad (86)$$

$$\frac{d\rho_{nk}}{dt} = i\frac{1}{\hbar}\rho_{nk}[V_{kk}(t) - V_{nn}(t)] - i\frac{1}{\hbar}(\rho_{kk} - \rho_{nn})V_{nk}(t)\exp(i\omega_{nk}t) - (\rho_{kk} - \rho_{nn})\gamma_{nk}\rho_{nk} \quad (87)$$

Assuming the function $\rho_{nn}$ varies slowly compared with the rapidly oscillating function $\rho_{nk}$, it is easy to find the solution of equation (87), treating the function $\rho_{nn}$, as a parameter. This solution has the form

$$\rho_{nk} = -i\frac{1}{\hbar}(\rho_{kk} - \rho_{nn})\exp[f(t)]\int_{-\infty}^{t}V_{nk}(t')\exp[i\omega_{nk}t' - f(t')]dt' \quad (88)$$

where

$$f(t) = -i\frac{1}{\hbar}\int[V_{nn}(t) - V_{kk}(t)]dt - (\rho_{kk} - \rho_{nn})\gamma_{nk}t \quad (89)$$

Substituting the expression (88) into (86), one obtains

$$\frac{d\rho_{nn}}{dt} = \frac{g}{\hbar^2}(\rho_{kk} - \rho_{nn}) - 2\gamma_{nk}\rho_{nn}\rho_{kk} \quad (90)$$

where

$$g = \int_{-\infty}^{t}\overline{V_{nk}(t)V_{nk}^*(t')\exp[i\omega_{nk}(t-t') + f^*(t) - f^*(t')]}dt' + \int_{-\infty}^{t}\overline{V_{nk}(t')V_{nk}^*(t)\exp[-i\omega_{nk}(t-t') + f(t) - f(t')]}dt' \quad (91)$$

This equation has been averaged over the fast oscillations in which the parameters $\rho_{nn}$ and $\rho_{kk}$ can be considered to be constant.

3.2 Thermostatistics of the excited states of the atom

In general, to find the excited state of an atom in interaction with a thermal reservoir, it is necessary to solve equations (83), (84). For a two-level atom these equations take the form (86), (87) or (90). To achieve a solution, it is necessary to know the functions $V_{nk}(t)$, which are usually unknown. Moreover, these functions are random, so any solution to these equations would describe only one of the possible implementations of the statistical process. For this reason, as usual, what are of interest are not the specific solutions of equations (83), (84), but their statistical properties. To determine these, one needs to know the statistical properties of the functions $V_{nk}(t)$, which are also unknown. However, the stationary states of an atom interacting with a thermal reservoir can be found from quite general considerations.

Let us proceed from the fact that the system should have one or more stationary states, at least one of which must be stable. It is this stable stationary state of the system that will be observed in experiments because random perturbations from thermal collisions lead to the fact that only a stable state of all of the possible stationary states will "survive". If the system has a stable stationary state, then the system should have a Lyapunov function $F(\rho_{11}, \rho_{22}, \rho_{33}, \dots)$, which depends on all of the parameters $\{\rho_{nn}\}$, and which will reflect the tendency of the system to the stable stationary state. To fix ideas, let us assume that in a stable stationary state of the system under consideration, the Lyapunov function has a maximum. Therefore, its time derivative must be non-negative, at least in the neighbourhood of the stable stationary state; i.e., $\frac{dF}{dt} \geq 0$, the equality corresponding to only the stationary state.

Such a Lyapunov function for the two-level atom can be built easily from general considerations. Let parameter $\rho_{nn}^{(0)}$ correspond to a stable stationary state. Next, the function $F_n$, satisfying the condition

$$\frac{dF_n}{dt} = -\frac{d\rho_{nn}}{dt} \ln \frac{\rho_{nn}}{\rho_{nn}^{(0)}} \tag{92}$$

will have all of the above-mentioned properties of the Lyapunov function.

In fact, $\frac{dF_n}{dt} = 0$ when $\frac{d\rho_{nn}}{dt} = 0$, a condition achieved only when $\rho_{nn} = \rho_{nn}^{(0)}$. Let us prove that in all other cases (i.e., when $\frac{d\rho_{nn}}{dt} \neq 0$) we will have $\frac{dF_n}{dt} > 0$. Indeed, if $\rho_{nn} < \rho_{nn}^{(0)}$ then the function $\rho_{nn}$ will grow, tending to the stationary state $\rho_{nn} = \rho_{nn}^{(0)}$; i.e., it will mean $\frac{d\rho_{nn}}{dt} > 0$, and, consequently, $\frac{dF_n}{dt} > 0$. If for some reason $\rho_{nn} > \rho_{nn}^{(0)}$, then, tending to the stationary state $\rho_{nn} = \rho_{nn}^{(0)}$, the function $\rho_{nn}$ will decrease: i.e., it will mean $\frac{d\rho_{nn}}{dt} < 0$, and therefore, again, $\frac{dF_n}{dt} > 0$.

Thus the function $F_n$ is always increasing, which means that in the steady state $\rho_{nn} = \rho_{nn}^{(0)}$, it has a maximum (at least locally).

Let $\rho_{kk}^{(0)}$ corresponds to the same stable stationary state of the two-level atom, i.e., $\rho_{nn}^{(0)} + \rho_{kk}^{(0)} = 1$. Then, similarly, it is easy to prove that the function $F_k$, satisfying the condition

$$\frac{dF_k}{dt} = -\frac{d\rho_{kk}}{dt} \ln \frac{\rho_{kk}}{\rho_{kk}^{(0)}} \tag{93}$$

will also be a monotonically increasing function of time, and in a stable stationary state it will have a maximum.

Obviously, the function

$$F = F_n + F_k \tag{94}$$

also possesses these properties because

$$\frac{dF}{dt} = -\frac{d\rho_{nn}}{dt} \ln \frac{\rho_{nn}}{\rho_{nn}^{(0)}} - \frac{d\rho_{kk}}{dt} \ln \frac{\rho_{kk}}{\rho_{kk}^{(0)}} \tag{95}$$

Using expression (24), the function

$$F = -\left(\rho_{nn} \ln \frac{\rho_{nn}}{\rho_{nn}^{(0)}} + \rho_{kk} \ln \frac{\rho_{kk}}{\rho_{kk}^{(0)}}\right) + \lambda_0(\rho_{nn} + \rho_{kk}) \tag{96}$$

will satisfy the same conditions, where $\lambda_0$ is an arbitrary constant. The function (96) will be the Lyapunov function for the system (86), (87).

The function (96) can also be rewritten in the form

$$F = -(\rho_{nn} \ln \rho_{nn} + \rho_{kk} \ln \rho_{kk}) + \left(\rho_{nn} \ln \rho_{nn}^{(0)} + \rho_{kk} \ln \rho_{kk}^{(0)}\right) + \lambda_0(\rho_{nn} + \rho_{kk}) \tag{97}$$

Generalizing function (97) for the case of an atom with an arbitrary number of excited eigenmodes, which in the thermal environment are described by equations (83) and (84), one can show that it has the Lyapunov function

$$F = -\sum_n \rho_{nn} \ln \rho_{nn} + \sum_n \rho_{nn} \ln \rho_{nn}^{(0)} + \lambda_0 \sum_n \rho_{nn} \tag{98}$$

In this case, using expression (85), one obtains

$$\frac{dF}{dt} = -\sum_n \frac{d\rho_{nn}}{dt} \ln \frac{\rho_{nn}}{\rho_{nn}^{(0)}} \tag{99}$$

Indeed, let each parameter $\rho_{nn}$ monotonically tends to its steady-state value $\rho_{nn}^{(0)}$. One can consider separately the functions $F_n$, satisfying the conditions $\frac{dF_n}{dt} = -\frac{d\rho_{nn}}{dt} \ln \frac{\rho_{nn}}{\rho_{nn}^{(0)}}$. By analogy with the above, it is easy to prove that the functions $F_n$ will increase monotonically with time and reach a maximum in a stable stationary state $\rho_{nn}^{(0)}$. Next, the function

$$F = \sum_n F_n \tag{100}$$

that satisfies condition (99) and, therefore, has the form (98), will be the Lyapunov function for the system under consideration.

Let us introduce the notations

$$S = -\sum_n \rho_{nn} \ln \rho_{nn} \tag{101}$$

$$E = -\frac{1}{\beta}\sum_n \rho_{nn} \ln \rho_{nn}^{(0)} - \frac{\mu}{\beta}\sum_n \rho_{nn} \tag{102}$$

where $\beta$ and $\mu$ are arbitrary constants.

Next, the function (98) can be rewritten in the form

$$F = S - \beta E + \lambda \sum_n \rho_{nn} \tag{103}$$

where $\lambda = \lambda_0 - \mu$.

The absolute maximum of the function $F$ corresponds to the stable stationary state of the system under consideration. This maximum is found from the ordinary extremum condition for a continuously differentiable function

$$\frac{\partial F}{\partial \rho_{nn}} = 0 \text{ for all } n \qquad (104)$$

At the same time, from expression (103), it is clear that the absolute maximum of the function $F$ can be interpreted as the conditional maximum of the function $S$ under given constraints: a given value of the parameter $E$ (102) and the normalization condition (85). Then, searching for the maximum of the function $F$ in the form (103) represents a standard method of undetermined Lagrange multipliers for finding the conditional maximum of the function $S$ under the given constraints (85) and (102). In this case, the Lyapunov function (103) will be the corresponding Lagrange function, while the parameters $\beta$ and $\lambda$ are the undetermined Lagrange multipliers.

Thus we came to the following conclusion: a stable stationary state of the system corresponds to the conditional maximum of the function (101) under the given constraints (85) and (102).

Being multiplied by the Boltzmann constant, the form of the function $S$ (101) coincides with the Boltzmann-Gibbs entropy for the "ensemble of the atom's eigenmodes"; however, this function will not be entropy in its statistical meaning because the parameters $\rho_{nn}$ are not probabilities, but represent the portion of the continuous electric charge of the electron wave in the eigenmode $n$ of an atom [5]. In other words, the parameters $\rho_{nn}$ are the real parameters that describe the state of a continuous deterministic electron wave in an atom, but not the probabilities of finding the atom in different discrete states, as it is usually interpreted in quantum mechanics.

The normalization condition (85) is the law of conservation of electric charge, and follows directly from the nonlinear Schrödinger equation (5), and, consequently, from equations (83), (84). Thus it is an integral of equations (83), (84).

The question remains: what is the parameter $E$ defined by expression (102)? Obviously, it should also be an integral of equations (83) and (84). The natural integral of the equations for the system under consideration is the energy of the atom, which is defined by the expression [5]

$$E = \sum_n \hbar \omega_n \rho_{nn} \qquad (105)$$

Then, using the method of undetermined Lagrange multipliers (103), (104), one obtains

$$\rho_{nn} = z^{-1} \exp\left(-\frac{\hbar \omega_n}{kT}\right) \qquad (106)$$

where as usual, the notation is introduced

$$\beta = \frac{1}{kT} \qquad (107)$$

in which $k$ is the Boltzmann constant, $T$ is the temperature, and $z$ is the constant determined by substituting expression (106) into the normalization condition (85):

$$z = \sum_n \exp\left(-\frac{\hbar\omega_n}{kT}\right) \tag{108}$$

Therefore, we have in fact obtained the usual Gibbs distribution, with the only difference that it describes not the probability of the atom remaining in the different discrete states, but the actual distribution of the electric charge of the classical continuous electron wave between the eigenmodes of the atom.

The purpose of such a detailed analysis, whose result, in fact, is the usual Gibbs distribution, is that in this case we are not dealing with discrete objects (particles or discrete states), for which one can enter the state probabilities associated with their energies, but with continuous classical waves. Therefore, the distribution (106) (108) describes the properties of the continuous classical electron wave in an atom excited by thermal collisions, and the functions (106) describe the intensities of the excitations of the atom's different eigenmodes having the eigenfrequencies $\omega_n$. In this case the parameter $\hbar\omega_n$ cannot be considered as the energy of an atom in a discrete state $n$ because the energy of the atom is described by expression (105) and is a continuous variable. Thus, the result obtained above is not trivial because the physical meaning of the distributions (106), (108) is fundamentally different from the physical meaning of the usual Gibbs distribution, which describes discrete objects or discrete states.

In particular, for the two-level atom, when only two eigenmodes $k$ and $n$ are excited, one obtains

$$\rho_{nn} = \frac{\exp\left(-\frac{\hbar\omega_{nk}}{kT}\right)}{1+\exp\left(-\frac{\hbar\omega_{nk}}{kT}\right)}, \quad \rho_{kk} = \frac{1}{1+\exp\left(-\frac{\hbar\omega_{nk}}{kT}\right)} \tag{109}$$

3.3 Thermal radiation

Knowing the thermostatistics of the excited states of an atom (106) (108), it is possible to calculate the intensity of spontaneous (electric dipole) emission (66) at a frequency $\omega_{nk}$:

$$I_{nk} = \frac{4\omega_{nk}^4 |\mathbf{d}_{nk}|^2}{3c^3} \frac{1}{\sum_s \exp\left(-\frac{\hbar\omega_{sn}}{kT}\right)} \frac{1}{\sum_s \exp\left(-\frac{\hbar\omega_{sk}}{kT}\right)} \tag{110}$$

Then, the full emission intensity of the atom is

$$I = \sum_n \sum_{k<n} I_{nk} \tag{111}$$

In particular, for the two-level atom, when only two eigenmodes $k$ and $n$ are excited, one obtains

$$I = I_{nk} = \frac{4\omega_{nk}^4 |\mathbf{d}_{nk}|^2}{3c^3} \frac{\exp\left(-\frac{\hbar\omega_{nk}}{kT}\right)}{\left[1+\exp\left(-\frac{\hbar\omega_{nk}}{kT}\right)\right]^2} \tag{112}$$

The spontaneous emission (110), (112) can be called thermal emission because it is caused by the excitation of the eigenmodes of the atom from its thermal collisions with the atoms of the thermal reservoir at temperature $T$.

Thus, an atom in the mixed state (106), (108) continuously emits energy in the form of electromagnetic waves with intensity (110), and thus permanently loses energy. Therefore it tends to arrive into a pure state in which only the lower of the eigenmodes is excited. To remain in the stationary state (106), (108), the atom must continuously receive energy from outside at the rate of its emission losses. This energy can be supplied as a result of thermal collisions with the atoms of the thermal reservoir, and of interaction with the radiation field. The case in which energy is supplied only through interaction with the radiation field was considered in Section 2.3.

In this section, let us consider the case when the atom undergoes only thermal effects, and the radiation field around it is absent (i.e., it is assumed that radiation is carried away in the surrounding space). Such an atom is a non-equilibrium thermodynamic system and it is necessary to continuously supply energy to it to compensate for losses from spontaneous (thermal) emission.

According to the expression (105), the energy of the two-level atom [5]

$$E = \hbar\omega_n \rho_{nn} + \hbar\omega_k \rho_{kk} \tag{113}$$

With expression (24), one obtains

$$\frac{dE}{dt} = \hbar\omega_{nk} \frac{d\rho_{nn}}{dt} \tag{114}$$

Substituting the expression (90) into (114), one obtains

$$\frac{dE}{dt} = \hbar\omega_{nk} \frac{g}{\hbar^2} (\rho_{kk} - \rho_{nn}) - 2\hbar\omega_{nk}\gamma_{nk}\rho_{nn}\rho_{kk} \tag{115}$$

The term

$$W_T = \hbar\omega_{nk} \frac{g}{\hbar^2} (\rho_{kk} - \rho_{nn}) \tag{116}$$

is the rate at which the atom exchanges energy with the thermal reservoir. The last term in the right-hand side of equation (115), according to (68), is the intensity of emission of the atom. Equation (115) can be written in the form of the energy balance equation:

$$\frac{dE}{dt} = W_T - I \tag{117}$$

Because in this case energy is supplied to the atom thermally, thermodynamics dictates that this is possible only if the temperature of the thermal reservoir is higher than the "temperature of the atom", as a thermodynamic system. In this case

$$W_T = \alpha(T - T_a) \tag{118}$$

where $\alpha$ is the coefficient of heat exchange between the atom as a thermodynamic system and the thermal reservoir, $T$ is the temperature of the thermal reservoir, $T_a$ is the "temperature of the atom". In equilibrium, the energy that the atom loses through emission is fully compensated by a resupply of thermal energy from the thermal reservoir, In this case $W_T = I$, i.e.,

$$\alpha(T - T_a) = I \tag{119}$$

It follows that the "temperature of the atom" as a thermodynamic system will be different from the temperature of the thermal reservoir:

$$T_a = T - I/\alpha \tag{120}$$

and it is always lower than the reservoir temperature, which supports the thermal emission. In this case, the energy will be continuously "pumped through" the atom as a thermodynamic system at the rate $I$. In other words, such an atom is an open thermodynamic system in a state of forced equilibrium.

It is obvious that precisely the temperature $T_a$ defined by expression (120) should enter into the distribution (106), (108).

Let us explain what "the temperature of the atom" means in this case. Atoms, interacting with the thermal reservoir, continuously and randomly change their states. However, all of the random states will be described by some statistics; in particular, the mean intensities of the excitations of the eigenmodes of an atom will be described by the distribution (106) (108). If we take many identical atoms that interact with the thermal reservoir but not with each other, then such a system will have the same statistics as a single atom, and can be considered as a "gas". Some temperature can be ascribed to such a system ("gas"). This is the temperature considered above to be the "temperature of the atom" $T_a$.

3.4 Joint action of the thermal collisions and the radiation field

Let us consider the case when the radiation generated by the atoms remains in the same region of space as the atoms and influences them. In this case, each atom experiences both thermal effects from its environment (thermal reservoir) and the effects of an external electromagnetic field generated by emission from other atoms (and itself).

Combining equations (20), (21) and (80), (81), one obtains

$$i\hbar \frac{dc_n}{dt} = c_k\{V_{nk}(t)\exp(i\omega_{nk}t) - \int_0^\infty (\mathbf{d}_{nk}\mathbf{E}_\omega)\exp[-i(\omega - \omega_{nk})t]\,d\omega\} + c_n V_{nn}(t) - i\hbar\gamma_{nk}c_n|c_k|^2 \tag{121}$$

$$i\hbar \frac{dc_k}{dt} = c_n\{V_{nk}^*(t)\exp(-i\omega_{nk}t) - \int_0^\infty (\mathbf{d}_{nk}^* \mathbf{E}_\omega^*)\exp[i(\omega - \omega_{nk})t]\,d\omega\} + c_k V_{kk}(t) +$$
$$i\hbar \gamma_{nk} c_k |c_n|^2 \quad (122)$$

Similarly, by combining equations (26), (27) and (86), (87), one obtains

$$\frac{d\rho_{nn}}{dt} =$$
$$i\frac{1}{\hbar}\rho_{nk}^*\{\int_0^\infty (\mathbf{d}_{nk}\mathbf{E}_\omega)\exp[-i(\omega-\omega_{nk})t]\,d\omega - V_{nk}(t)\exp(i\omega_{nk}t)\} -$$
$$i\frac{1}{\hbar}\rho_{nk}\{\int_0^\infty (\mathbf{d}_{nk}^*\mathbf{E}_\omega^*)\exp[i(\omega-\omega_{nk})t]\,d\omega - V_{nk}^*(t)\exp(-i\omega_{nk}t)\} - 2\gamma_{nk}\rho_{nn}\rho_{kk} \quad (123)$$
$$\frac{d\rho_{nk}}{dt} = i\frac{1}{\hbar}\rho_{nk}[V_{kk}(t) - V_{nn}(t)] + (\rho_{kk} - \rho_{nn})\{i\frac{1}{\hbar}\int_0^\infty (\mathbf{d}_{nk}\mathbf{E}_\omega)\exp[-i(\omega-\omega_{nk})t]\,d\omega -$$
$$i\frac{1}{\hbar}V_{nk}(t)\exp(i\omega_{nk}t) - \gamma_{nk}\rho_{nk}\} \quad (124)$$

As above, assuming the function $\rho_{nn}$ to be slowly varying compared with the rapidly oscillating function $\rho_{nk}$, one obtains the solution of equation (124) in the form

$$\rho_{nk} = i\frac{1}{\hbar}(\rho_{kk} - \rho_{nn})\exp[f(t)]\int_{-\infty}^t \{\int_0^\infty (\mathbf{d}_{nk}\mathbf{E}_\omega)\exp[-i(\omega-\omega_{nk})t' - f(t')]\,d\omega -$$
$$V_{nk}(t')\exp(i\omega_{nk}t' - f(t'))\}dt' \quad (125)$$

where

$$f(t) = i\frac{1}{\hbar}\int[V_{kk}(t) - V_{nn}(t)]dt - (\rho_{kk} - \rho_{nn})\gamma_{nk}t \quad (126)$$

Let us substitute the function (125) into equation (123), and average it over the fast oscillations (both of the radiation field, and those related to the thermal collisions), during which the parameter $\rho_{nn}$ can be considered constant. Doing this, we should take into account that because of the random natures of the radiation field and the thermal collisions, they do not correlate. Executing transforms similar to those executed in sections 2.1 and 3.1, one obtains

$$\frac{d\rho_{nn}}{dt} = \frac{4\pi}{3\hbar^2}|\mathbf{d}_{nk}|^2 \int_0^\infty \frac{(\rho_{kk}-\rho_{nn})^2 \gamma_{nk}}{(\omega-\omega_{nk})^2 + (\rho_{kk}-\rho_{nn})^2 \gamma_{nk}^2} U_\omega(\omega)d\omega + \frac{g}{\hbar^2}(\rho_{kk} - \rho_{nn}) - 2\gamma_{nk}\rho_{nn}\rho_{kk} \quad (127)$$

where the parameter $g$ is defined by expression (91).

In particular, for the radiation field, which has a wide spectrum, taking into account the results of Section 2.2, one obtains

$$\frac{d\rho_{nn}}{dt} = (\rho_{kk} - \rho_{nn})\left[\frac{g}{\hbar^2} + \frac{4\pi^2|\mathbf{d}_{nk}|^2}{3\hbar^2}U_\omega(\omega_{nk})\right] - 2\gamma_{nk}\rho_{nn}\rho_{kk} \quad (128)$$

Substituting the derivative (128) into the expression (114), and using (68), (72) and (116), one can write the energy balance equation for the atom under study in the form

$$\frac{dE}{dt} = W_T + W_E - I \quad (129)$$

If atom is in equilibrium, then $\frac{dE}{dt} = 0$ and

$$W_T + W_E - I = 0 \quad (130)$$

Using (118), this equilibrium condition can be written in the form

$$\alpha(T - T_a) + W_E - I = 0 \qquad (131)$$

If, at the same time, the atom is in thermodynamic equilibrium with the thermal reservoir, then according to the thermodynamics, $T = T_a$. Next, it follows from the equation (131) that the atom is also in detailed equilibrium with the radiation field: $W_E = I$. In this case, the loss of energy of the atom due to emission will be compensated by the work of the electromagnetic radiation field (see. Section 2.3).

## 4  Equilibrium Thermal Radiation

### 4.1  Temperature of the radiation field

A certain temperature $T_r$ can be attributed to thermal radiation as a thermodynamic system. The temperature of a system of discrete particles is an understandable and commonly experienced parameter and is related to the distribution of these particles over their possible energy levels (discrete or continuous). For this reason, in quantum mechanics, which considers the radiation as a system of particles (photons), the temperature of radiation field has a simple and intuitive classical meaning.

Considering the radiation as a classical electromagnetic field, the question is, how can a temperature be attributed to such a continuous system? For this purpose one can use the thermodynamic condition of equilibrium, according to which two systems in thermodynamic equilibrium must have the same temperature. Thus, if the radiation is in equilibrium with the thermal reservoir, according to the thermodynamics it can be said to have a temperature equal to that of the thermal reservoir: $T_r = T$. It is this radiation we call the equilibrium thermal radiation, in contrast to the non-equilibrium thermal radiation, which is also caused by the thermal excitations of the atoms, but which is not in equilibrium with them (see. Section 3.3).

Let us consider now the case when there are three systems: (i) the thermal reservoir, (ii) the thermal radiation and (iii) an atom, and they are all in thermal equilibrium. According to thermodynamics, if the system A is in thermodynamic equilibrium with system B, and the system B is in thermodynamic equilibrium with the system C, then the systems A and C are also in thermodynamic equilibrium. It follows that in equilibrium, the thermodynamic temperature of the thermal reservoir, of an atom and of the thermal radiation are the same:

$$T_r = T_a = T$$

From this condition and from the equilibrium condition (131) follows expression (58), which indicates that in the system consisting of the thermal reservoir, an atom and the thermal radiation being in thermodynamic equilibrium, the detailed balance of the atom with the radiation field described by the condition (58), and the detailed balance of the atom with the thermal reservoir, described by the distribution (106) (108), must exist simultaneously. It should be noted that this conclusion is not trivial, taking into account that the atom is under the simultaneous effects of the radiation field and thermal excitation from the thermal reservoir.

4.2 Spectral energy density of the equilibrium thermal radiation

As shown above, a two-level atom, being in a mixed excited state characterised by the parameters $\rho_{nn}$ and $\rho_{kk}$ connected by the condition (24), will be in equilibrium with the radiation field only when the spectral energy density of the radiation $U_\omega(\omega)$ satisfies the condition (74). Note that in deriving the equilibrium conditions (74), the reason for an atom being in an excited state was not specified, so the expression (74) is valid for any cause of the excitation of the eigenmodes of the atom.

In particular, if the two-level atom is in equilibrium with the thermal reservoir, it will be in mixed excited state described by the parameters (109).

Substituting (109) into expression (74) one obtains an expression for the spectral energy density of equilibrium thermal radiation

$$U_\omega(\omega) = \frac{\hbar\omega^3}{\pi^2 c^3} \frac{1}{\left[\exp\left(\frac{\hbar\omega}{kT}\right)-1\right]\left[1+\exp\left(-\frac{\hbar\omega}{kT}\right)\right]} \tag{132}$$

This expression can be rewritten in the equivalent form

$$U_\omega(\omega) = \frac{\hbar\omega^3}{2\pi^2 c^3 \sinh\left(\frac{\hbar\omega}{kT}\right)} \tag{133}$$

where $\sinh x = \frac{1}{2}(e^x - e^{-x})$ is the hyperbolic sine.

Expression (132) is different from Planck's law for equilibrium thermal radiation by the additional factor $1/\left[1 + \exp\left(-\frac{\hbar\omega}{kT}\right)\right]$.

Figure 1 shows a comparison of expression (132) with Planck's law. It can be observed that the difference associated with the factor $1/\left[1 + \exp\left(-\frac{\hbar\omega}{kT}\right)\right]$ is small and noticeable only at low frequencies $\frac{\hbar\omega}{kT} < 4$.

It is interesting to compare expression (132) with experimental data [8], which is an experimental basis of the theory of thermal radiation. Such a comparison is shown in Fig. 2 for the spectral function $U_\lambda = \frac{\omega^2}{2\pi c} U_\omega$.

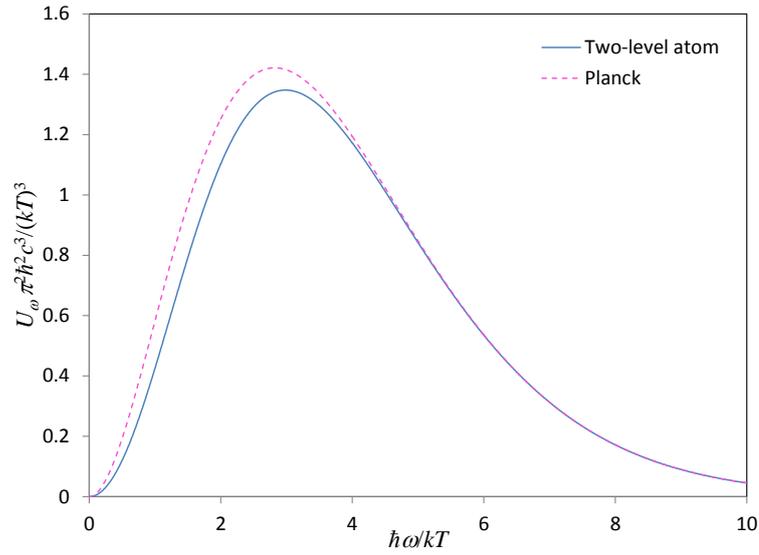

Fig. 1. Dependence of the non-dimensional spectral energy density of equilibrium thermal radiation on the non-dimensional frequency, calculated by using Planck's law and the model of the two-level atom (132).

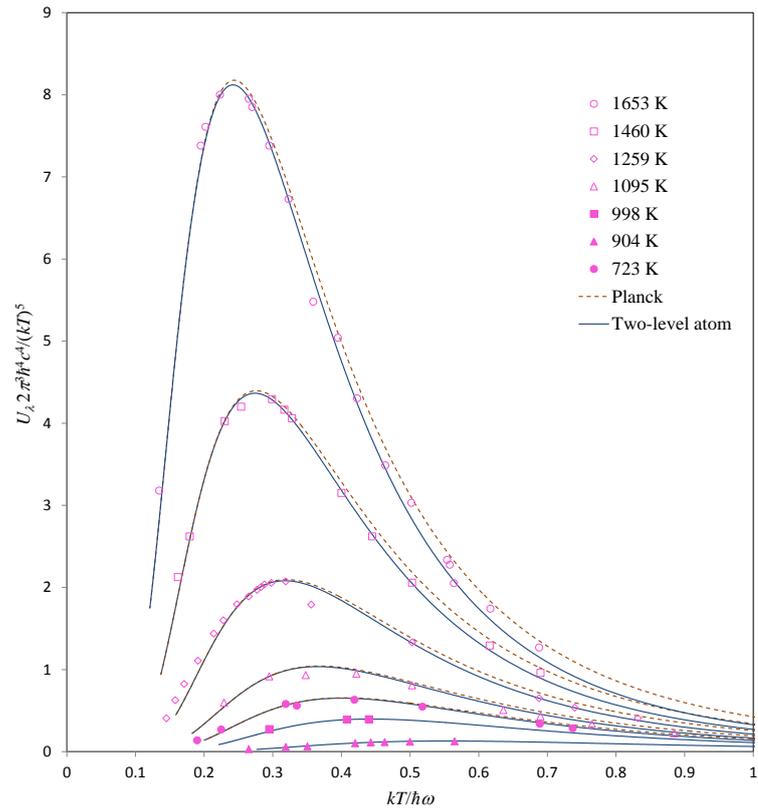

Fig. 2. Comparison of the non-dimensional spectral energy density $U_\lambda$ calculated with the model of the two-level atom (132) (solid line) and the experimental data [8] (markers) for different temperatures. The dashed lines show the dependencies calculated by Planck's law.

Taking into account the experimental errors (including temperature measurement error), it can be said that both expression (132) and Planck's law describe the experimental data [8] equally well. The difference between the results of calculations with expression (132) for the two-level atom and Planck's law is so small that it appears difficult to determine which describes the experimental data better.

At the same time, we note that expression (132) for the spectral energy density of equilibrium thermal radiation is derived fully within the framework of classical field theory without using the hypothesis of energy quantization. From the standpoint of the proposed theory, Planck's law is approximate, and should be replaced (for the two-level atom, at least) by the more accurate expression (132).

It is easy to see that for the spectral energy density of radiation (132), the Wien displacement law $\lambda_{max} = b/T$ and the Stefan-Boltzmann law $u = \sigma T^4$ retain their forms; however, the constants within them should be slightly corrected compared with those calculated by Planck's law.

After a simple calculation using expression (132), one obtains

$$\left(\frac{\hbar\omega}{kT}\right)_{max} = 3$$

instead of $\left(\frac{\hbar\omega}{kT}\right)_{max} = 2.821439$ based on Planck's law. Accordingly, the Wien constant obtained by the expression (132) is $b = 0.003081$ m·K instead of $b = 0.002898$ m·K as obtained by Planck's law.

Similarly, for the Stefan-Boltzmann constant, using formula (132), one obtains (see. Appendix)

$$\sigma = \frac{\pi^2 k^4}{16\hbar^3 c^3}$$

instead of $\sigma = \frac{\pi^2 k^4}{15\hbar^3 c^3}$ obtained by Planck's law.

4.3 Einstein A-coefficient for spontaneous emission

Let us consider the equation (47), which describes the interaction between an atom and a broad-spectrum radiation field. This equation can be rewritten in the form

$$\frac{d\rho_{nn}}{dt} = -\frac{d\rho_{kk}}{dt} = B_{kn}U_\omega(\omega_{nk})\rho_{kk} - B_{kn}U_\omega(\omega_{nk})\rho_{nn} - A_{nk}\rho_{nn}\rho_{kk} \qquad (134)$$

where

$$B_{kn} = B_{nk} = \frac{4\pi^2|\mathbf{d}_{nk}|^2}{3\hbar^2} \qquad (135)$$

$$A_{nk} = 2\gamma_{nk} \qquad (136)$$

Using expression (28), one can find the connection between the coefficients (135) and (136)

$$B_{kn} = B_{nk} = \frac{\pi^2 c^3}{\hbar \omega_{nk}^3} A_{nk} \qquad (137)$$

Note that this connection is universal because it does not depend on the electric dipole moments $|\mathbf{d}_{nk}|^2$ and is determined only by the frequency of spontaneous emission $\omega_{nk}$. Therefore, although the relation (137) was obtained for the hydrogen atom, one can expect that it will be valid for other atoms having different frequencies of spontaneous emission $\omega_{nk}$.

Once again note that all of the results above have been obtained within the framework of classical field theory without any quantization.

However, it is possible to interpret equation (134) in the spirit of a photon hypothesis. Let us introduce the notations

$$w_{kn}^{ind} = B_{kn} U_\omega(\omega_{nk}), \quad w_{nk}^{ind} = B_{nk} U_\omega(\omega_{nk}), \quad w_{nk}^{sp} = A_{nk} \rho_{kk} \qquad (138)$$

Next, the equation (134) can be written in the form

$$\frac{d\rho_{nn}}{dt} = -\frac{d\rho_{kk}}{dt} = w_{kn}^{ind} \rho_{kk} - w_{nk}^{ind} \rho_{nn} - w_{nk}^{sp} \rho_{nn} \qquad (139)$$

This equation, formally, has the form of the kinetic equation describing the transitions of some fictitious system between two discrete states $n$ and $k$; in this case, taking condition (24) into account, the parameters $\rho_{nn}$ and $\rho_{kk}$ can be interpreted as the probabilities of finding the system in the corresponding discrete states. The first and second terms on the right-hand side of equation (139) can be interpreted as induced transitions $k \to n$ and $n \to k$, respectively, and the third term as a spontaneous transition $n \to k$. Accordingly, the parameters $w_{kn}^{ind}$ and $w_{nk}^{ind}$, which are determined by the expressions (138), should be interpreted as the probabilities of the corresponding induced transitions (related to the absorption and stimulated emission of a "photon" $\hbar \omega_{nk}$) per unit time, and the parameter $w_{nk}^{sp}$, as the probability of spontaneous transition (associated with the spontaneous emission of a "photon" $\hbar \omega_{nk}$) per unit time. In this case, the parameters $B_{kn}$, $B_{nk}$ and $A_{nk}$ (135), (136) should be interpreted as the Einstein coefficients. Taking expression (137) into account, the "probabilities" of the corresponding "transitions" are connected by the relations

$$w_{kn}^{ind} = w_{nk}^{ind} = \frac{\pi^2 c^3}{\hbar \omega_{nk}^3} \frac{1}{\rho_{kk}} w_{nk}^{sp} U_\omega(\omega_{nk}) \qquad (140)$$

These expressions are different from those are derived in quantum electrodynamics [9] only by the additional factor $\frac{1}{\rho_{kk}}$.

In the case when the excitation of the upper eigenmode $n$ is weak, i.e., when $\rho_{nn} \ll \rho_{kk}$, one can take $\rho_{kk} \approx 1$, and then relation (140) becomes the well-known result of quantum electrodynamics [9]. In this case the correct expressions (137) for the Einstein A-coefficient for spontaneous emission are obtained, and expression (132) for the spectral energy density of equilibrium thermal radiation becomes Planck's law.

However, the above analysis shows that the photon interpretation of the equations obtained is not only unnecessary but also erroneous because it is based only on the outer analogy. We see that the expression for the spectral energy density of equilibrium thermal radiation (132) and the so-called Einstein A-coefficient for spontaneous emission can be obtained in a natural way within the framework of the classical field theory without any quantization of radiation.

Moreover, we see that Planck's radiation law and the Einstein's statistical interpretation of equation (134) correspond only to the weak excitation of the upper mode of the electron wave in an atom when $\rho_{nn} \ll \rho_{kk}$. If the excitation of the upper modes in the atom is not weak, it is necessary to use the nonlinear equation (134), and for the spectral energy density of equilibrium thermal radiation, it is necessary to use not Planck's law but a more complicated expression, e.g., in the case of the two-level atom, expression (132).

## 5 Concluding Remarks

Thus, in this study, for the first time, we have managed to build a completely classical theory of thermal radiation using only classical ideas about light as a continuous classical electromagnetic wave and to obtain Planck's spectrum of thermal radiation.

From the above analysis, it follows that in contrast to the predictions of Planck's law, the spectral energy density of thermal radiation is apparently not a universal function of frequency but depends, albeit slightly, on the properties of the emitting atoms. This can be observed by considering, for example, the three-level atom. Only in the case of a weak excitation of the atom can one consider that thermal radiation from all types of sources (atoms) is universal and is described by the unified Planck's law.

Most of the results obtained above refer to equilibrium thermal radiation. However, obviously, there are processes in which the thermal radiation from atoms (i.e., the radiation induced by the thermal excitation of the mixed states of the atoms) will not be in equilibrium with them, and hence will not be equilibrium thermal radiation. This is possible, for example, in cases when the electromagnetic waves emitted by the atoms are carried away in space and do not affect other atoms of the radiating system (an open system).

The intensity of arbitrary (not necessarily equilibrium) thermal radiation is described by the expression (112). The frequency corresponding to the maximum of the intensity of the radiation (112) at a constant parameter $|\mathbf{d}_{nk}|^2$, is determined by the solution of the transcendental equation

$$x \frac{1 - \exp(-x)}{1 + \exp(-x)} = 4$$

where $x = \frac{\hbar \omega}{kT}$. The solution of this equation is $x \approx 4.13$. In reality, however, to determine this maximum it is necessary to take into account the change of the parameter $|\mathbf{d}_{nk}|^2$ with frequency $\omega_{nk}$. Nevertheless, this indicates that the frequency at which a maximum of the intensity of the non-equilibrium thermal radiation is observed can be different from the frequency corresponding to the maximum of the intensity of equilibrium thermal radiation as determined by Planck's law or by the expression (132).

From here, one can make an important practical conclusion. The temperatures of many objects that cannot be measured directly (e.g., extremely high temperatures or the temperatures of remote objects, including astrophysical objects) is determined indirectly using their thermal radiation and Wien's displacement law, assuming that the radiation is in equilibrium with the radiating atoms. If this is not so, then the use of the Wien's constant obtained from the Planck's law, or even from the expression (132), can result in significant error in determining the temperature of a radiating object.

The results of this paper and papers [5,6] show that there is no point in dividing the emission of atoms into spontaneous emission and stimulated emission: any emission of an atom is conventional classical electric dipole radiation [5] that occurs when an atom is in a mixed excited state for any reason.

In conclusion, note that the quantum mythology is based on some basic physical effects that, as considered, cannot be explained within the framework of classical ideas and require quantization. These effects include: (i) thermal radiation, (ii) the photoelectric effect, (iii) the Compton effect, (iv) discrete events observed in the interaction of light with matter (detector), primarily the double-slit experiments with a weak light source, (v) electron diffraction, (vi) the properties of atoms, primarily their stability and their discrete emission spectra. Summarizing the interim results of this series of papers (which includes this paper and papers [1-6]), it can be argued that all these effects have a quite simple and, most importantly, clear explanation within the framework of classical field theory, if we assume that the electron wave and electromagnetic waves are continuous classical fields.

Other so-called "quantum" effects will be examined from the point of view of classical field theory in the following papers of this series.

**Acknowledgments**

Funding was provided by Tomsk State University competitiveness improvement program.

# Appendix

Using the spectral energy density of radiation (132), the Stefan-Boltzmann constant is defined by the expression

$$\sigma = \frac{k^4}{\hbar^3 \pi^2 c^3} \int_0^\infty \frac{x^3}{[\exp(x)-1][1+\exp(-x)]} dx \qquad (A1)$$

Let us calculate the integral in the expression (A1). Obviously,

$$\int_0^\infty \frac{x^3}{[\exp(x)-1][1+\exp(-x)]} dx = \int_0^\infty \frac{x^3 \exp(-x)}{[1-\exp(-2x)]} dx = \int_0^\infty x^3 \sum_{n=0}^\infty e^{-(2n+1)x} dx =$$

$$\sum_{n=0}^\infty \frac{1}{(2n+1)^4} \int_0^\infty y^3 e^{-y} dy = \frac{\pi^4}{16} \qquad (A2)$$

because the last integral in expression (A2) is equal to 6, and

$$\sum_{n=0}^\infty \frac{1}{(2n+1)^{2p}} = \frac{2^{2p}-1}{2 \cdot (2p)!} B_p$$

where $p = 1,2,3,...$; $B_p$ are the Bernoulli numbers; $B_2 = 1/30$.

# References

1. Rashkovskiy S.A. Quantum mechanics without quanta: the nature of the wave-particle duality of light // Quantum Studies: Mathematics and Foundations, (Published online 31 October 2015), DOI: 10.1007/s40509-015-0063-5; see also arXiv 1507.02113 [quant-ph] (PART 1).
2. Rashkovskiy S. A. Are there photons in fact? Proc. SPIE. 9570, The Nature of Light: What are Photons? VI, 95700G. (September 10, 2015) doi: 10.1117/12.2185010.
3. Rashkovskiy S.A. Semiclassical simulation of the double-slit experiments with single photons. Progress of Theoretical and Experimental Physics, 2015 (12): 123A03 (16 pages) DOI: 10.1093/ptep/ptv162.
4. Rashkovskiy S.A. Quantum mechanics without quanta. arXiv 1507.02113 [quant-ph] (PART 2: The nature of the electron), 2016, 61p.
5. Rashkovskiy S.A. Classical theory of the hydrogen atom, arXiv:1602.04090 [physics.gen-ph], 2016, 32 p.
6. Rashkovskiy S.A. Classical-field description of the quantum effects in the light-atom interaction, arXiv:1603.02102 [physics.gen-ph], 2016, 28 p.
7. L.D. Landau, E.M. Lifshitz. *The Classical Theory of Fields*. Vol. 2 (4th ed.). Butterworth-Heinemann (1975).
8. Lummer, O., & Pringsheim, E. Kritisches zur schwarzen Strahlung. *Annalen der Physik*, *311*(9), 192-210 (1901).

9. V.B. Berestetskii, E.M. Lifshitz, L.P. Pitaevskii. *Quantum Electrodynamics*. Vol. 4 (2nd ed.). Butterworth-Heinemann (1982).